\documentclass[preprint,superscriptaddress,showpacs,amsfonts,amsmath,floatfix]{revtex4}

\usepackage{graphicx}
\usepackage{epstopdf}
\usepackage{amssymb}
\usepackage{amsmath}
\usepackage{hyperref}
\usepackage{xspace}
\usepackage{color}

\newcommand{\bra}[1]{\left\langle #1 \right|}
\newcommand{\ket}[1]{\left| #1 \right\rangle}
\newcommand{\aket}[1]{\left| \phi_{#1} \right\rangle}

\begin{document}

\title {Thermal effects on the sudden changes and freezing of correlations between remote atoms in cavity QED network} 

\author{Vitalie Eremeev\footnote{veremeev@fis.puc.cl}}
\affiliation{Facultad de Ingenier\'ia, Universidad Diego Portales, Santiago, Chile}

\author{ Nellu Ciobanu}
\affiliation{Facultad de F\'isica, Pontificia Universidad Cat\'olica de Chile, Casilla 306, Santiago, Chile}

\author{ Miguel Orszag}
\affiliation{Facultad de F\'isica, Pontificia Universidad Cat\'olica de Chile, Casilla 306, Santiago, Chile}

\begin{abstract}
We investigate the thermal effects on the sudden changes and freezing of the quantum and classical correlations of remote qubits in a cavity quantum electrodynamics (CQED) network with losses. We find that the detrimental effect of the thermal reservoirs on the system can be compensated via an efficient coupling of the fiber connecting the two cavities of the system. Furthermore, for certain initial conditions, we find a double sudden transitions in the dynamics of Bures geometrical quantum discord. The second transition tends to disappear at a critical temperature, hence freezing the discord. Finally, we discuss some ideas of the experimental realization of the present proposal.
\end{abstract}


 \maketitle
 
\noindent 
In recent years, quantum correlations have played a fundamental role in quantum computation and quantum information processing  \cite{Nielsen}.
However, Entanglement, a popular measure of such correlations, is not the only type and other more general correlations such as the quantum discord \cite{{Vedral2001}, {Zurek2001}}, geometric quantum discord as Bures distance \cite{{Spehner}, {Aaronson2013}}, \textit{etc.} have been shown to capture, even in completely separable systems, non-classical behavior such as computational speed-up in quantum information processes \cite{Datta2008}.
The unusual dynamics of the classical and quantum decoherence originally reported in \cite{{Maziero2009}, {Mazzola2010}} and confirmed experimentally in \cite{{Xu}, {Auccaise2011}}, stimulated a very high interest in the investigation of the phenomena of sudden changes in the correlations for different physical systems, like in the optical or condensed matter domain. Hence, during the last four years, intensive efforts were focused in order to explain fundamentally the nature of the sudden transition and freezing effects of the quantum correlations and under which conditions such transitions occur. Also, from the perspective of the applications, how efficiently one could engineer these phenomena in quantum information technologies.
As was observed and concluded in the majority of the studies, the puzzling peculiarities of the sudden transitions and freezing phenomena \cite{{Maziero2009}, {Auccaise2011}, {Xu}, {Pinto2013}, {He2013}, {Mazzola2010}, {Lo Franco2012}, {You2012}, {Aaronson2013}} are hidden in the structure of the density operator during the whole evolution of a bipartite quantum open system. Since the first works \cite{{Maziero2009}, {Mazzola2010}} the intriguing question remains still open - how these fascinating effects are affected by the presence of the noisy environments and if there are efficient mechanisms to control them in both non dissipative or dissipative decoherence models. The sudden transitions and freezing effects of the classical and quantum correlations were observed in several different systems with different kind of decoherence mechanisms \cite{{Maziero2009}, {Auccaise2011}, {Xu}, {Pinto2013}, {He2013}, {Mazzola2010}, {Lo Franco2012}, {You2012}, {Aaronson2013}}, but to the best of our knowledge there is modest state-of-the-art  research about the influence of the real thermal environments on the correlations for quantum-optical systems such as CQED networks.  Beginning with the first conceptual ideas \cite {net1997} the theoretical and experimental studies advanced in time \cite{Serafini2006} until some very recent studies \cite {net}.

\begin{figure} [t]
\includegraphics[width=7.7 cm]{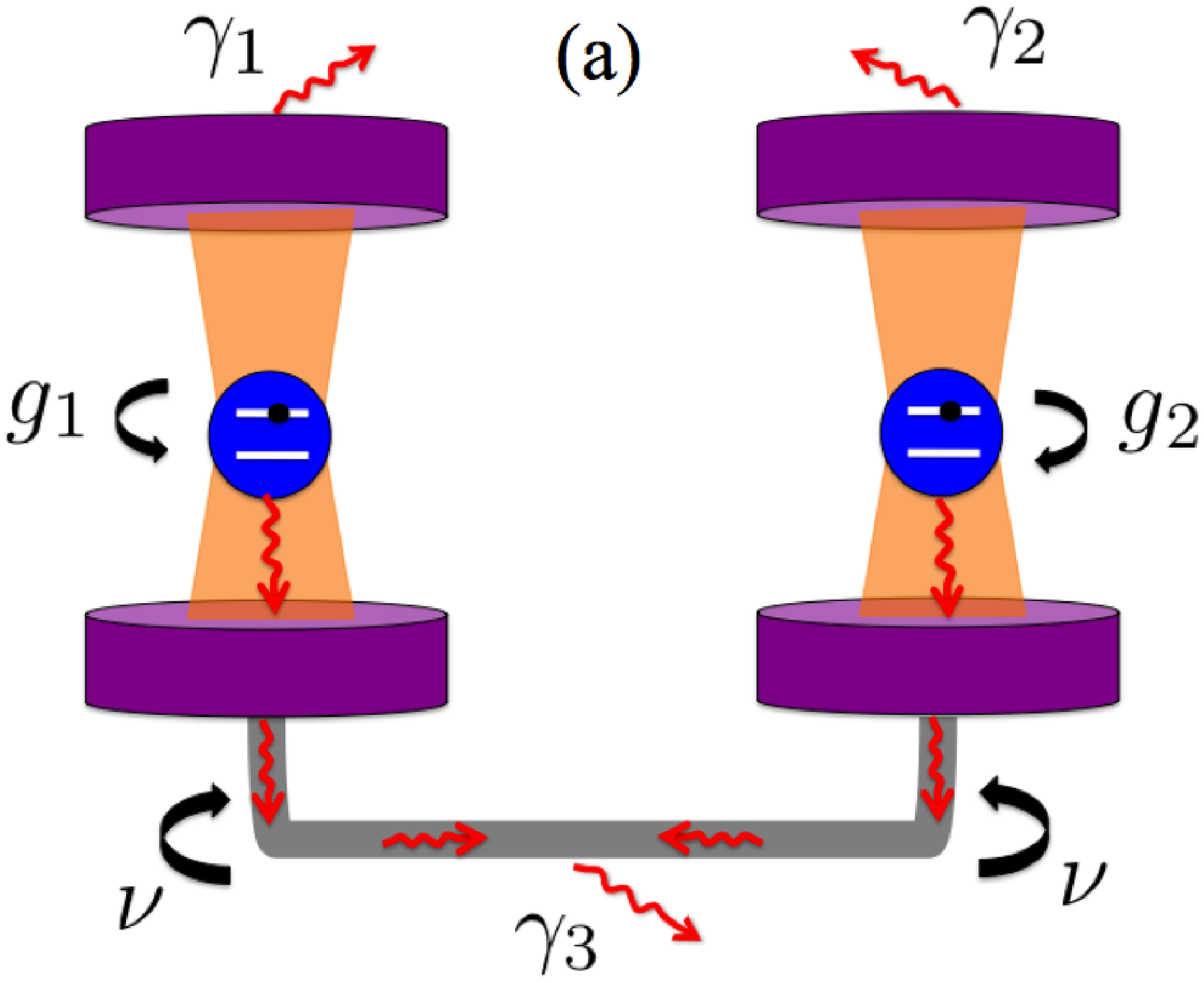}
\includegraphics[width=8.6 cm]{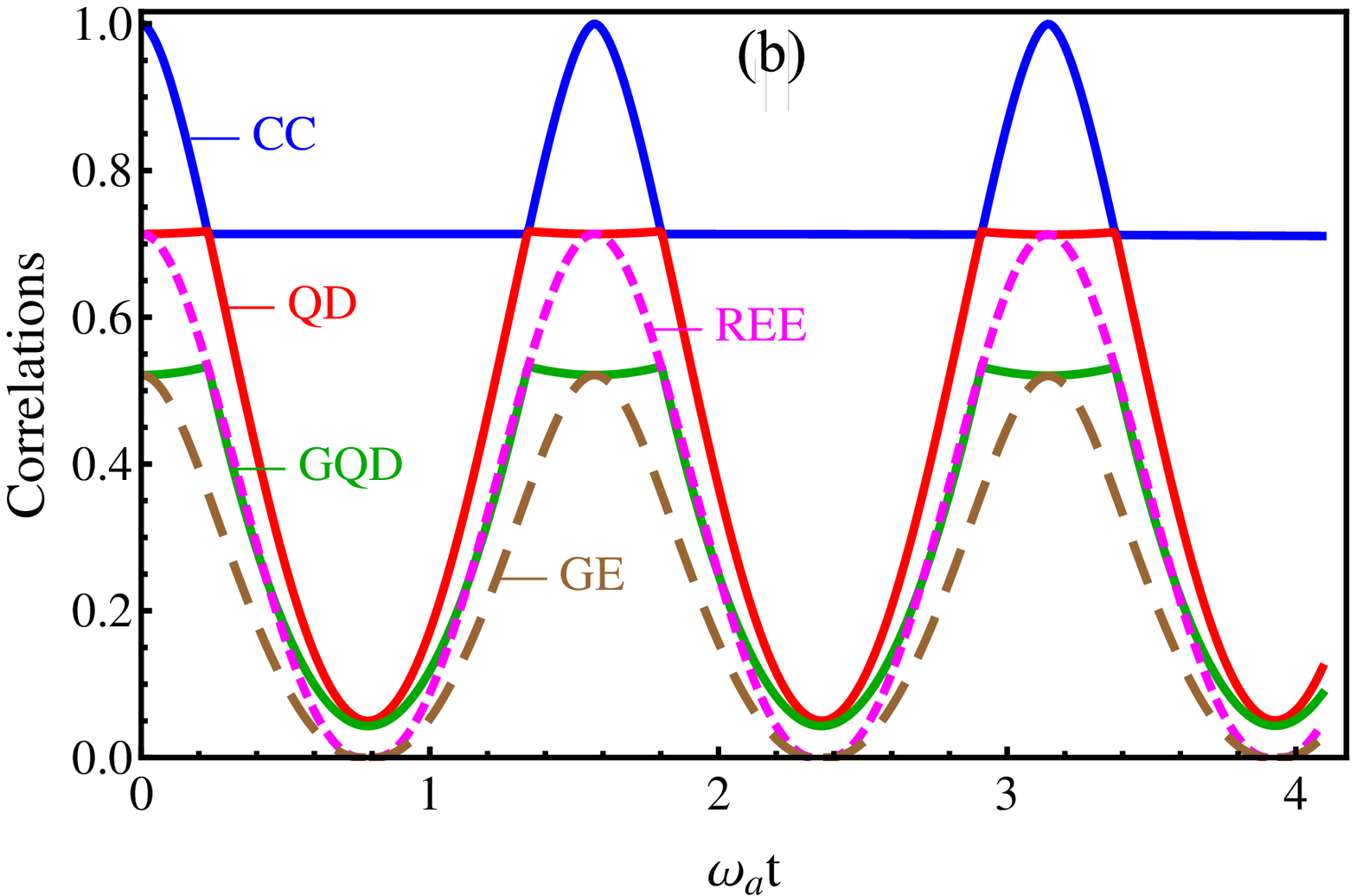}
\caption{(a) Remote atoms in CQED network. (b) Time evolution of the correlations: CC (blue solid),  QD (red solid), GQD (green solid), REE (magenta dashed) and GE (brown long dashed) for the reservoirs at zero temperatures. The parameters are the same as in Fig. \ref{fig2}(a).} 
\label{fig1}
\end{figure}

In  Fig. \ref{fig1}(b) we show the time evolution of the classical and quantum correlations for an optical network discussed in \cite {Eremeev2012} and applied to the case of two excitations with the qbits initially prepared in a class of states as in Eq. (\ref{rhoBD}) for all the reservoirs at zero temperature. We observe the quantum-classical sudden transition in case of our CQED model similar to other studied systems \cite{{Maziero2009}, {Auccaise2011}, {Xu}, {Pinto2013}, {He2013}, {Mazzola2010}, {You2012}, {Lo Franco2012}, {Aaronson2013}}. Besides the classical correlations (CC), entropic quantum discord (QD) and relative entropy of entanglement (REE), we have also studied two geometrical measures, the geometric entanglement (GE) and geometric quantum discord (GQD) defined with Bures distance \cite{Spehner}. We notice that the Bures GQD and QD show a similar behavior, having flat regions not affected by the dissipation processes during a particular time period, effect known as freezing of QD. At the same time CC decay and meet QD in a point where a sudden change occurs. After this point the CC remain constant during another time period until other sudden change follows and so we observe periodic revival of the correlations. On the other hand, the quantum correlations measured by the entanglement show different dynamics, evidencing effects of sudden death and birth not appearing in the QD and GQD for the given system.  

In the following we present briefly the model proposed in \cite{{Montenegro2011}, {Eremeev2012}} schematically shown in Fig.\ref{fig1}(a) and recall the analytical equations for a generalized model presented in this Letter. Basically the model considers a quantum open system of two remote qubits (two-level atoms) where each qubit interacts independently with one of the distant cavities coupled by a transmission line (e.g., fiber). For the sake of simplicity one considers the approach of short fiber limit: only one active mode of the fiber interacts with the cavity modes \cite{Serafini2006}. The whole system is open because the cavities and fiber exchange energy with their individual thermal environments, hence we have a very general case of dissipative decoherence of the quantum system.
The Hamiltonian of the composite system under the rotating-wave approximation (RWA) in units of $\hbar$ reads
\begin{eqnarray}
H_s &=& \omega_f a_3^{\dag} a_3+\sum_{j=1}^2  \left( \omega_a S_{j,z}+\omega_0 a^{\dag}_j a_j \right) + \sum_{j=1}^2 \left( g_j S^+_j a_j + \nu a_3 a^{\dag}_j + H.c.\right),
\label{Ham}
\end{eqnarray}
where $a_1(a_2)$ and $a_3$ is the boson operator for the cavity 1(2) and the fiber mode, respectively;
 $\omega_0$, $\omega_f$  and $\omega_a$ are the cavity, fiber and atomic frequencies, respectively;  $g_j$($\nu$) the atom(fiber)-cavity coupling constants; $S_{z}$, $S^{\pm}$ are the atomic inversion and ladder operators, respectively.

One of the important advance and novelty of the present model compared to the previous ones \cite{{Montenegro2011}, {Eremeev2012}} is based on the generalization to large number of excitations in the system. To the best of our knowledge, this approach of many excitations in similar systems \cite{{net1997}, {Serafini2006}} is not common, and may be one of few existent studies. 

To describe the evolution of an open quantum-optical system usually the approach of the \textit{Kossakowski-Lindblad} phenomenological master equation is considered with the system Hamiltonian decomposed on the eigenstates of the field-free subsystems. However, sometimes a CQED system is much more realistically modeled based on the microscopic master equation (MME), developed in \cite{Scala2007, Breuer} where the system-reservoir interactions are described by a master equation with the system Hamiltonian mapped on the atom-field eigenstates, known as dressed states. The present system consists in two atoms within their respective cavities and a fiber connecting them. We represent the leakage of the two cavities and the fiber via a coupling to individual external environments.
Thus, we identify three dissipation channels corresponding to each element of the system with its own reservoir. Commonly, in CQED the main sources of dissipation originate from the leakage of the cavity photons due to the imperfect reflectivity of the cavity mirrors. Another mechanism of dissipation corresponds to the spontaneous emission of photons by the atom, however this kind of loss is negligible small in the CQED regime considered in our model, and consequently is neglected. Hence, it is straightforward to bring the Hamiltonian $H_s$ in \eqref{Ham} to a matrix representation in the atom-field eigenstates basis. To define a general state of the whole system we use the notation: $\ket{i}=\ket{A_1}\otimes\ket{A_2}\otimes\ket{C_1}\otimes\ket{C_2}\otimes\ket{F} \equiv \ket{A_1A_2C_1C_2F}$, where 
${A}_{j=1,2}$ correspond to the atomic states, that can be $e(g)$ for excited(ground) state, while ${C}_{j=1,2}$ and ${F}$ define the cavities and fiber states, respectively, which may correspond to $0$, $1$, ... $n$ photon states. Because the quantum system is dissipative, the excitations may leak to the reservoirs degrees of freedom, hence the ground state of the system, $\ket{0}=\ket{gg000}$, should be also considered in the basis of the states. Therefore, in the case of $N$ excitations in our system, the number of dressed states, $\ket{i}$, having minimum one excitation, i.e excluding the ground state $\ket{0}$, is computed by a simple relation: $d_N=N+2\sum_{k=1}^N k(k+1)$. For example, in case of $N=2$ excitations the Hamiltonian $H_s$ in Eq. (\ref{Ham}) is decomposed in a state-basis of the dimension $1+d_2$, i.e. is a  $19\times19$ matrix; for 6 excitations $H_s$ is represented  by a $231\times231$ matrix, and so on. Hence it is evident that for large $N$ the general problem becomes hard to solve even numerically. In the present work, we develop our calculations up to 6 excitations which is an improvement as compared to some previous works, e.g. with two excitations \cite{Serafini2006}.

Considering the above assumptions and following the approach of \cite{Scala2007,Breuer}, the MME for the reduced density operator $\rho(t)$ of the system is defined in the form as given by Eq. (2) in \cite {Eremeev2012}. 
In the following we develop the equation for the density operator $\rho(t)$ mapped on the eigenstates basis, $\bra{\phi_m}\rho(t)\ket{\phi_n} = \rho_{mn}$ for the case of $N$ excitations in the system
\begin{eqnarray}
\dot{\rho}_{mn}&=& - i \bar{\omega}_{n,m} \rho_{mn} + \sum_{k=1}^{d_N} \big[ \frac{\gamma_{k \to 0}}{2} \big( 2\delta_{m0}\delta_{0n} \rho_{kk} - \delta_{mk}\rho_{kn} - \delta_{kn} \rho_{mk} \big)  \nonumber \\ 
&&+\frac{\gamma_{0 \to k}}{2} \big( 2\delta_{mk}\delta_{kn}  \rho_{00} - \delta_{m0}\rho_{0n} - \delta_{0n}\rho_{m0}  \big) \big],
 \label{rhosys}
\end{eqnarray}
here $\delta_{mn}$ is the Kronecker delta; the physical meaning of the damping coefficients $\gamma_{k \to 0}$ and $\gamma_{0 \to k}$ refer to the rates of the transitions between the eigenfrequencies $\Omega_k$ and $\Omega_0$ downward and upward, respectively, defined as follows $\gamma_{k \to 0}=\sum_{j=1}^3 c_i^2\gamma_j(\bar{\omega}_{0,k}) \left[\langle n(\bar{\omega}_{0,k})\rangle_{T_j} + 1\right] $ and by the Kubo-Martin-Schwinger condition we have $\gamma_j(-\bar{\omega})=\mathrm{exp}\left(-\bar{\omega}/ T_j\right) \gamma_j({\bar{\omega}})$, where $c_i$ are the elements of the transformation matrix from the states  $\{\ket{0}, \ket{1}, ... , \ket{d_N}\}$ to the states $\{\aket{0}, \aket{1}, ... , \aket{d_N} \}$ (similar to Eq. (14) and Appendix A in \cite{Montenegro2011}). 
Here $\langle n(\bar{\omega}_{\alpha, \beta})\rangle_{T_j} = \left ( \mathrm{e}^{(\Omega_\beta- \Omega_\alpha) / T_j} - 1\right )^{-1}$ corresponds to the average number of the thermal photons (with $k_B=1$). The damping coefficients play a very important role in our model because their dependence on the reservoirs temperatures imply a complex exchange mechanism between the elements of the system and the baths. Further, one solves numerically the coupled system of the first-order differential equations (\ref{rhosys}) and compute the evolution of different kind of correlations between the two distant atoms, given some finite temperature of the reservoirs. In order to get the reduced density matrix for the atoms one performs a measurement on the cavities and the fiber vacuum states, $\ket{000} = \ket{0}_{C1}\otimes \ket{0}_{C2} \otimes \ket{0}_{F}$, latter we will explain how this task can be realized experimentally. We find that, after the projection, the reduced atomic density matrix has a X-form and the correlations can be computed as developed in \cite{{Luo2008}, {Fanchini2010}, {Spehner}}. 
\begin{figure} [t]
\includegraphics[width=8 cm]{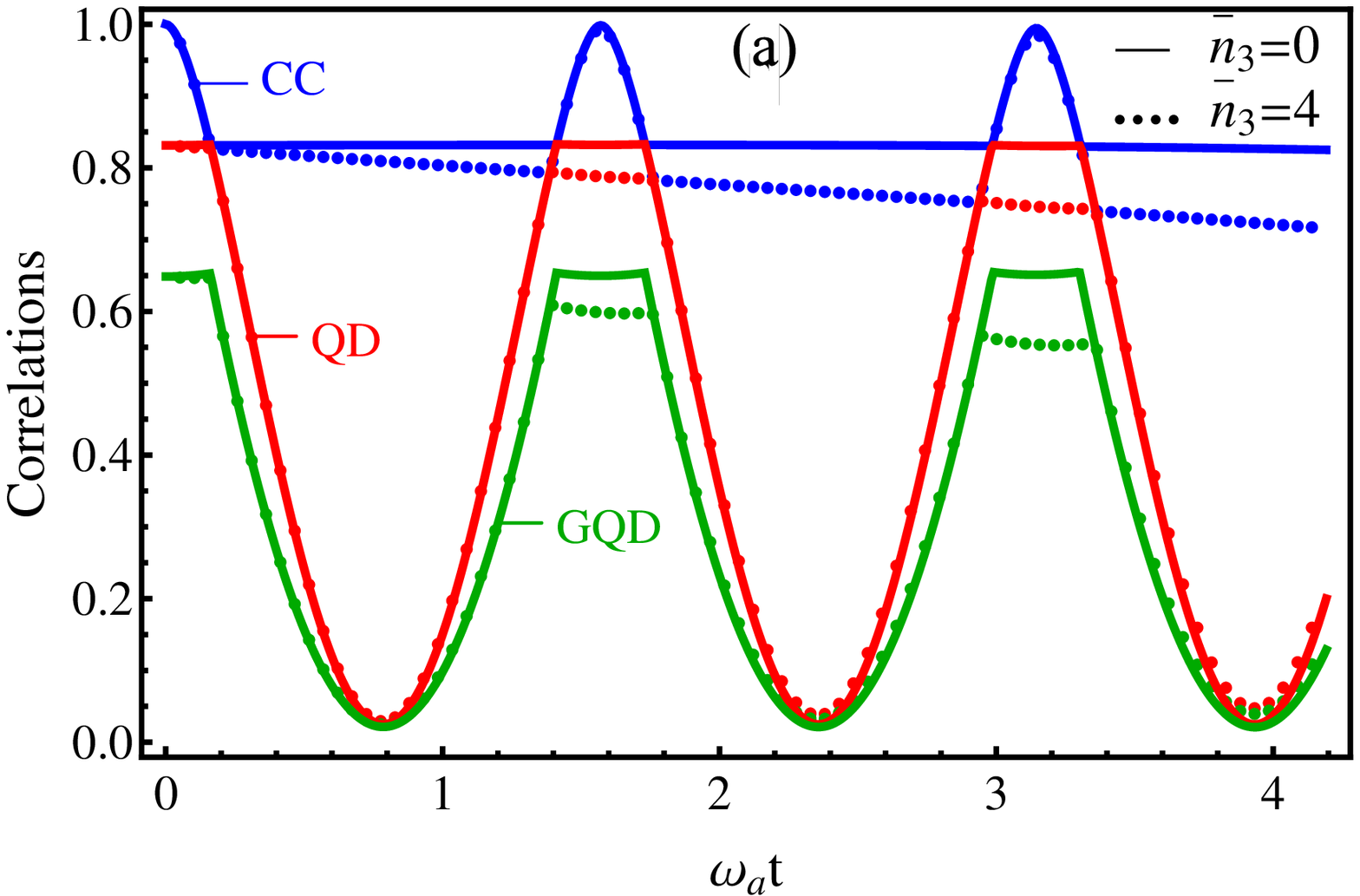}
\includegraphics[width=8 cm]{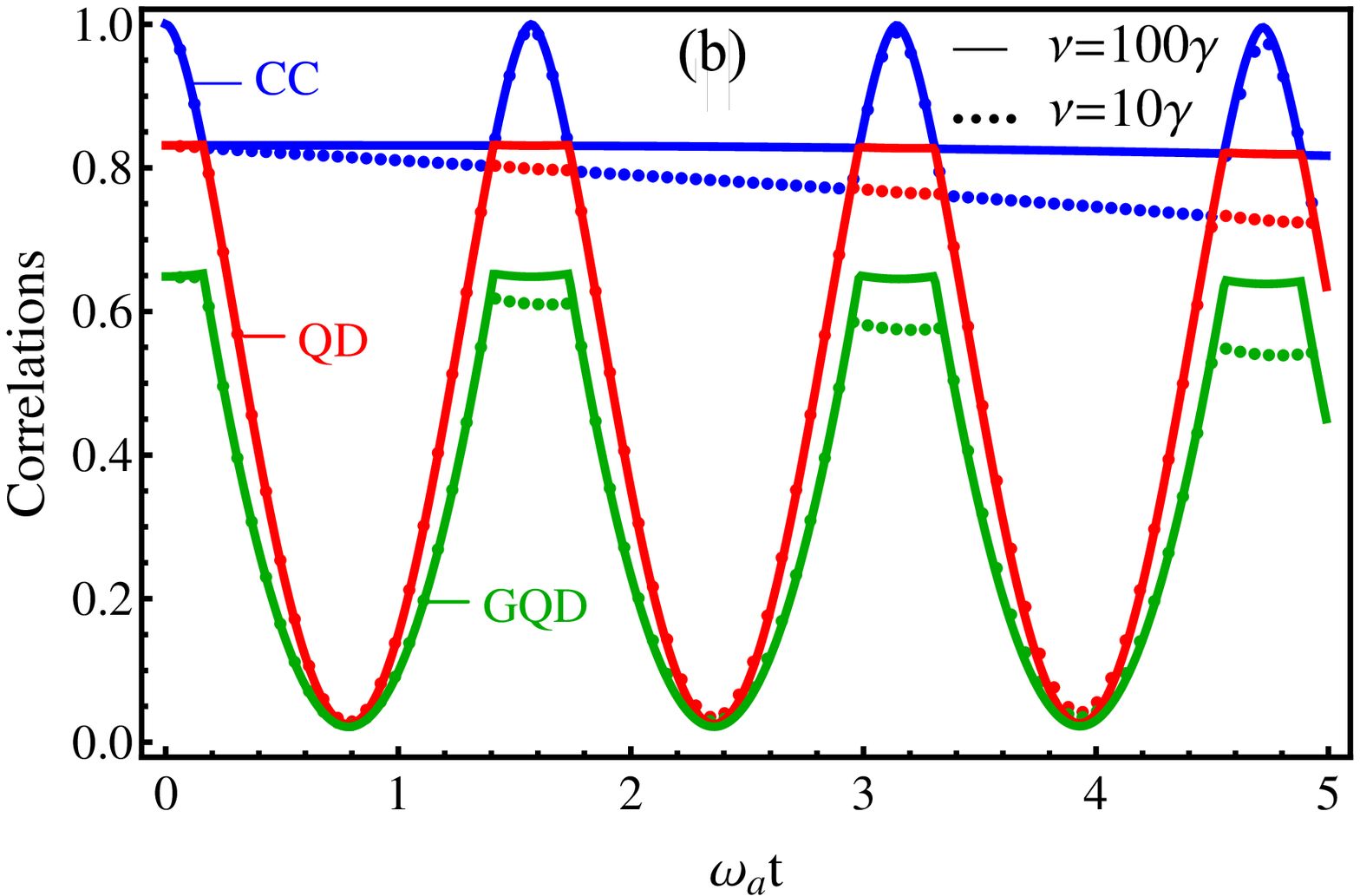}
\caption{The dynamics of the correlations: CC (blue), QD (red) and Bures GQD (green) for $\gamma=0.008 \omega_a$ and (a) varying the temperature of the fiber's reservoir given by the average number of the thermal photons, i.e. $\bar{n}_3=0$ (solid line) and $\bar{n}_3=4$ (dotted line) for constant cavity-fiber coupling $\nu=g=10\gamma$; (b) varying the cavity-fiber coupling, $\nu=10\gamma$ (dotted line) and $\nu=100\gamma$ (solid line) for constant $\bar{n}_3=3$.  The initial state is defined by $\vec{c}=(1, -0.95, 0.95)$ in Eq. (\ref{rhoBD}).}
\label{fig2}
\end{figure}

The system under consideration refers to the atoms with long radiative lifetimes trapped in their own cavities and connected by a fiber. The cavities and fiber exchange their energy with individual reservoirs Fig. \ref{fig1}(a) which, for the sake of simplicity, are taken to have the same damping rate $\gamma_1=\gamma_2=\gamma_3 \equiv \gamma$. The transition frequency of the atom is considered a free parameter used to scale the rest of the dimensionless parameters, which we take similar to some experimental data \cite{Raimond2001}, e.g. $\omega_a/2 \pi=10$ GHz. The atom-cavity couplings satisfy the constraint of the MME in a Markovian environment, i.e. $2g \gg \gamma$  \cite{Scala2007} and we set the values $g_1=g_2=g=10\gamma$ in the Figs.  \ref{fig1}-\ref{fig2}. The values of $\gamma$ and $\nu$ will be tuned to evidence the effects of the thermal baths. We find that the detunings haven't an important impact on the effects we discuss here, so setting the values $\omega_f=\omega_a$ and $\omega_a-\omega_0=0.1\omega_a$.
In the following, we compute the time evolution of the atomic correlations keeping in mind the main objective of this Letter, that is to find the influence of the thermal baths on these correlations. In order to compute the general correlations - classical and quantum for the given system, we consider the concepts of mutual information, classical correlations and entropic quantum discord as defined and calculated in \cite{{Vedral2001}, {Zurek2001}, {Luo2008}, {Fanchini2010}}, as well the geometric quantum discord with Bures distance, recently developed by one of us \cite{Spehner} and independently in \cite{Aaronson2013}. Let us consider that the two atoms initially are prepared in a particular state as Bell-diagonal (BD), described by a X-type density matrix 
in Bloch form as 
\begin{equation}
\rho(0)=[I\otimes I+\vec{c} \cdot (\vec{\sigma} \otimes \vec{\sigma})]/4,
 \label{rhoBD}
\end{equation}
where $\vec{\sigma}=(\sigma_1, \sigma_2, \sigma_3)$ is the vector given by Pauli matrices, $I$ is the identity matrix, and the vector $\vec{c}=(c_1, c_2, c_3)$ defines completely the state with $-1 \le c_i \le1$.  It is very important to point out here that the majority of the works, studying the sudden transition and freezing effects of the correlations in the bipartite quantum system, consider the special decoherence mechanisms (noise channels known as bit flip, phase flip and bit-phase flip \cite{{Mazzola2010}, {Aaronson2013}}) so that during the time evolution the density matrix preserves the property of its initial state, i.e. $\rho_{11}(t)=\rho_{44}(t)=[1+c_3(t)]/4$ and $\rho_{22}(t)=\rho_{33}(t)=[1-c_3(t)]/4$. For this scenario the classical correlations and quantum discord are easily computed with the help of the work of Luo in \cite{Luo2008} or given explicitly in \cite{{Mazzola2010}}. However, for quantum systems embedded in natural environments, like heat baths, the above  mentioned equality of the density matrix elements is no longer satisfied, as the system evolves in time, even if initially we have a BD state. Hence, it is of great interest to study the phenomena of sudden transition and freezing of the correlations for  more general and realistic dissipation models, such as our system, under the MME approach.

\begin{figure} [t]
\centering\includegraphics[width=12 cm]{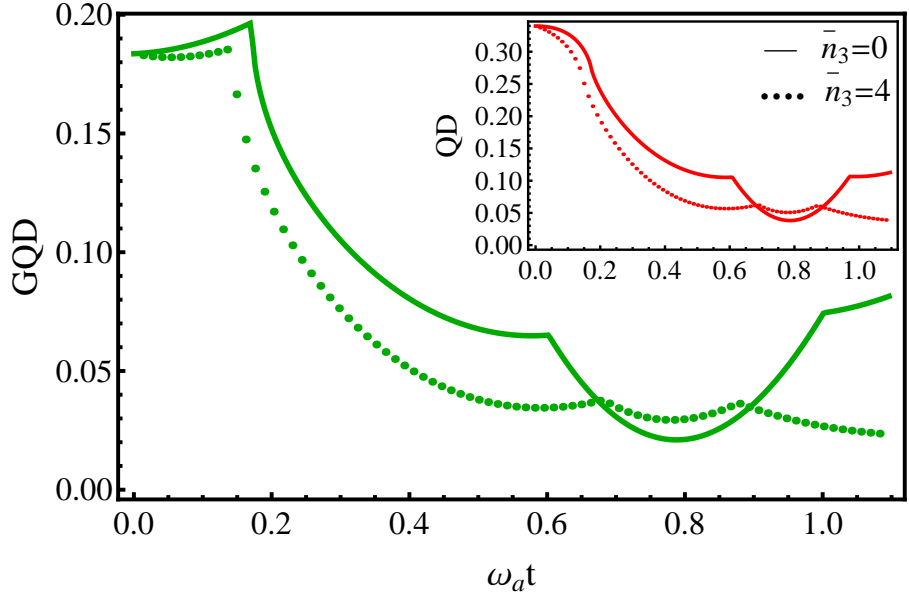}
\caption{Dynamics of GQD evidencing double sudden transitions (QD in the inset). The parameters considered here are $\gamma=0.1 \omega_a$, $\nu=g_1=g_2=5\gamma$ and $\bar{n}_3=0$ (solid line) and $\bar{n}_3=4$ (dotted line). The initial state is given by $\vec{c}=(0.85, -0.6, 0.36)$.}
\label{fig3}
\end{figure}

Recently, Pinto \textit{et al.} in \cite{Pinto2013} discussed the sensitivity of the sudden change of the QD to different initial conditions. In this context, the present work shed more light on this important subject. In particular, for the system under study, we find that the atomic density matrix preserves the initial BD form only for a short time under the action of the heat reservoirs. It very quickly evolves into a more general X-shaped non-BD form. Under these circumstances, we compute the quantum discord as shown in Figs. \ref{fig2}-\ref{fig3} by using a more general algorithm developed in \cite{Fanchini2010}. Furthermore, we also make use of an alternative, non-entropic measure of the quantum correlations such as the geometric quantum discord (GQD) with Bures distance, proposed and calculated for a bipartite system in \cite{Spehner}. Hence, observing in Fig. \ref{fig1}(b) the effect of the sudden change and freezing of the correlations for the reservoirs at zero temperature, it is natural to inquire about the thermal effects on the classical and quantum correlations when the cavities and fiber are connected to reservoirs at finite (non-zero) temperatures. In our numerical analysis we find that the freezing effects of the QD and GQD decay by increasing individually or collectively the temperatures of the cavities or the fiber. In the  Fig. \ref{fig2}(a) we show the effect of heating the fiber to four thermal photons and observe that the thermal effects act destructively on the freezing of both the entropic and geometric discords. However, the sudden transitions persist. The next question is: Could one recover from the damaging effects of the system being coupled to the thermal reservoirs? We find that we could in principle engineer such a recovery by a suitable increase in the fiber-cavity coupling. As a matter of fact, we show in Fig. \ref{fig2}(b) that, when we set the fiber's bath temperature to three thermal excitations, such recovery of the correlations, via the increase of the fiber-cavity coupling, is feasible, hence by this effect we understand the important role of the photon as the carrier of the quantum correlations between the remote qbits in such a network. 

Recently in \cite {Montealegre2013} the authors theoretically described another interesting class of sudden transitions and freezing of the quantum correlations, which later was observed experimentally in NMR setups \cite {Paula2013}. They found the formation of an environment induced double transition of Schatten one-norm geometric quantum correlations (GQD-1), which is not observed in the classical correlations, thus a truly quantum effect.
   
Motivated by this very recent findings we simulate the dynamics of the Bures GQD for our model and find a type of double sudden transitions somewhat different from the ones observed in \cite {{Montealegre2013}, {Paula2013}}. To the best of our knowledge, this kind of double sudden transitions and freezing effect for Bures GQD was not reported in literature and by this result we come to an important conclusion that the both, GQD-1 and Bures GQD evidence similar quantum effects. In Fig. \ref{fig3} we see the double sudden changes in the dynamics of Bures GQD for the reservoirs at zero temperatures, meanwhile the QD suffers one sudden change. 
We observe the following interesting result: as we increase the temperature of the fiber's bath, there is a peculiar tendency to freeze the GQD and the second transition tends to disappear, at a critical temperature. 

The experimental realization of the present proposal hinges on the possibility of realizing a quantum non-demolition (QND) measurements of the photon states in the fiber coupled cavities. There is an extensive literature on QND measurements in CQED, for a review see \cite {Grangier1998}. In our scheme we propose to measure the two-qubit density matrix in the condition that all the fields are in vacuum state, so it is feasible to monitor the probability of this state during the temporal evolution of the system \cite {Eremeev2012}. 

In summary, we analyzed here the phenomena of sudden changes and freezing of correlations in a CQED network with thermal dissipation channels (Fig. \ref{fig1}). Double sudden transitions for Bures GQD are observed for the first time. We conclude that by controlling the dissipation mechanisms one may engineer the quantum correlations with multiples sudden changes and freezing periods in the temporal evolution, effects which can find the practical applications. 
A kind of thermal critical effects in this model are expected like in other systems \cite {Werlang2010}.

\bigskip
We thank V\'ictor Montenegro for his valuable help with the numerical simulations. We acknowledge financial support from the Chilean Fondecyt project no.100039, and the project Conicyt-PIA anillo no. ACT-1112 'Red de an\'alisis estoc\'astico y aplicaciones'.

\end{document}